# Teaching introductory STEM with the Marble Game

Peter Hugo Nelson

Physics and Biology, Benedictine University, Lisle IL 60532
September 6, 2012; Revised October 21, 2012

Recently there have been multiple calls for curricular reforms to develop new pathways to the science, technology, engineering and math (STEM) disciplines. The Marble Game answers these calls by providing a conceptual framework for quantitative scientific modeling skills useful across all the STEM disciplines. The approach actively engages students in a process of directed scientific discovery. In a "Student Assessment of their Learning Gains" (SALG) survey, students identified this approach as producing "great gains" in their understanding of real world problems and scientific research. Using the marble game, students build a conceptual framework that applies directly to random molecular-level processes in biology such as diffusion and interfacial transport. It is also isomorphic with a reversible first-order chemical reaction providing conceptual preparation for chemical kinetics. The computational and mathematical framework can also be applied to investigate the predictions of quantitative physics models ranging from Newtonian mechanics through RLC circuits. To test this approach, students were asked to derive a *novel* theory of osmosis. The test results confirm that they were able to successfully apply the conceptual framework to a new situation under final exam conditions. The marble game thus provides a pathway to the STEM disciplines that includes quantitative biology concepts in the undergraduate curriculum - from the very first class.

## INTRODUCTION

In the past decade there has been a growing movement calling for more quantitative content in the life sciences curriculum. The *Bio2010* report called for more mathematics, physical and information sciences to be taught to new biology students (NRC, 2003). The *Vision and Change* report (AAAS, 2011) identified 6 core competencies required for all students, including the abilities to: 1) apply the process of science; 2) use quantitative reasoning; 3) use modeling and simulation; 4) tap into the interdisciplinary nature of science; 5) communicate and collaborate with other disciplines; and 6) to understand the relationship between science and society. The recent report by the President's Council of Advisors on Science and Technology (PCAST) goes further and calls for widespread curricular reform to develop new pathways to the science, technology, engineering and math (STEM) disciplines (PCAST, 2012). This call has very recently been reinforced by a call for a "learning progression" for life science majors through the curriculum (Klymkowsky and Cooper, 2012).

The "Marble Game" (Figure 1) provides a new starting point for introducing quantitative content into the life sciences curriculum (Nelson, 2011b).[1] As outlined below, the marble game provides students with a conceptual framework that supports quantitative reasoning throughout the STEM disciplines. The marble game can be used in physics courses before Newtonian mechanics, providing a simpler (and more biologically relevant) entry-level quantitative model. In this paper, this approach will be presented along with evidence from upper division courses *Biophysics* and *Physiological Modeling*, but it is argued that these materials are accessible at the introductory level.

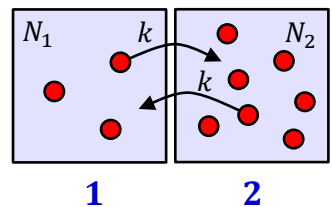

Figure 1. Schematic representation of the Marble Game. $N_1$ is the number of marbles in box 1, $N_2$ is the number of marbles in box 2, and there are a total of $N = N_1 + N_2$ marbles in the game. The figure shows an $N = 10$ marble game with $N_1 = 3$ marbles in box 1 and $N_2 = 7$ marbles in box 2. Marbles jump between boxes with rate constant $k$.

Physics is about quantitative scientific modeling. The basic idea is to develop a model of the thing that you are interested in and then to investigate the network of predictions that the model makes by comparing them with real-world data in a quantitative manner. In traditional physics instruction this begins with kinematics and projectile motion – but the scientific process (of investigating the predictions of kinematics) is usually ignored in traditional lecture classes. While kinematics might seem like a good place to start with physics (a ball thrown in the air is something we can all readily picture and are familiar with), there is a significant pedagogical problem. Many students do not have the required conceptual framework to understand "motion" properly at a mathematical level. Position, displacement, velocity and acceleration must all be understood correctly to appreciate the physics of the problem. The distinctions between these four vector quantities should be well understood by physics instructors, but students often find it extremely difficult to distinguish properly between them, particularly when they still have trouble with the concept of assigning a letter to represent a physical variable.

---

[1] © Peter Hugo Nelson 2012 pete@circle4.com
http://circle4.com/biophysics

Peter Hugo Nelson

In this paper, a different starting point for modeling physical systems is proposed. The marble game has numerous advantages over the traditional starting point (kinematics), the most important of which is that students do not have any prior experience with it. Hence, they have no preconceptions that need to be corrected. Another advantage is that only one variable must be understood – $N_1$ the number of marbles in box 1. It is very intuitive, being a simple count of the number of marbles in box 1. In addition, the marble game provides a base model for a learning progression (Klymkowsky and Cooper, 2012; Schwarz et al., 2009) that encompasses all of the STEM disciplines.

There is a growing movement to transform undergraduate education from a (sometimes boring) passive lecture-based process into an active-learning process based on educational research (DiCarlo, 2009). Basically the recommendation is that "we should teach the way we learn". Students should realize that science produces evidence-based knowledge and understanding – not just a list of declarations by some authority that need to be accepted and memorized without any evidence. A recent promo for the Science Channel sums up this idea in two words – "**question everything**" http://youtube.com/watch?v=IH5SQEKIGhA.

As discussed below, the marble game includes many of the features identified by multiple inter- and multi-disciplinary reports as being desirable (AAAS, 2011; Henderson and Dancy, 2009; HHMI-AAMC, 2009; NRC, 2003; NRC, 2011; NRC, 2012). From a disciplinary point of view, the marble game *directly* illustrates at least five of the seven general physiological models applicable from the molecular to organismal level (Modell, 2000). It also provides a prototypical model for molecular kinetic processes in biology, chemistry and physics that can be immediately applied to most biomedical engineering problems (Truskey et al., 2009). This stochastic framework is useful for modeling a vast array of processes including: nuclear decay, single-molecule dynamics, chemical kinetics, Brownian motion and diffusion, membrane transport for organelles, cells, organs, entire organisms and even evolution of populations and ecosystems. The marble game is a prototypical kinetic Monte Carlo (kMC) method that can be used to solve the master equations for stochastic processes (Markov chains). The marble game is based on kMC simulation methods that were developed independently for molecular transport processes (Nelson et al., 1991), but are isomorphic with methods developed earlier for solving the chemical master equation (CME) (Gillespie, 1977). These methods are gaining widespread adoption as a computational research technique for solving the CME. According to Beard and Qian "*We suggest that the importance of the CME to small biochemical reaction systems is on a par with the Boltzmann equation for gases and the Navier-Stokes equation for fluids. This is a big claim…*"(Beard and Qian, 2008). The marble game allows this technique to be introduced at the beginning of the STEM curriculum.

After students have learned how to play the marble game and apply it to realistic situations such as drug elimination, they then learn how to predict the "ensemble average" behavior of kMC sims using finite difference (FD) methods. These computational methods can be used in an introductory pedagogical setting instead of calculus. The marble game thus provides a mathematical and computational framework that can be applied to problems throughout the STEM disciplines (PCAST, 2012).

The marble game addresses all of the goals outlined in recent NRC reports on discipline-based education research (NRC, 2011; NRC, 2012):

- master a few major concepts well and in-depth;
- retain what is learned over the long-term;
- build a mental framework that serves as a foundation for future learning;
- develop visualization competence including ability to critique, interpret, construct, and connect with physical systems;
- develop skills (analytic and critical judgment) needed to use scientific information to make informed decisions;
- understand the nature of science; and
- find satisfaction in engaging in real-world issues that require knowledge of science.

The marble game addresses these goals by providing a mental model that can be used as a foundation for a quantitative framework that spans the STEM disciples. Students gain in-depth knowledge of a few important systems by deliberately practicing the scientific method (Deslauriers et al., 2011). The pedagogical approach relies heavily on interpretation and critical evaluation of student-generated graphs. The connection with the physical system is reinforced with questions and follow-up comments in the form of "About what you discovered" sections. By developing quantitative models and then critically comparing them with experimental data, students gain in-depth experience with applying the scientific method to real world problems (circle4.com/biophysics/modules). Student evaluations of these materials (Results) indicate that they have learned both metacognitive and procedural skills after working with the modules, gaining a realistic understanding of what is possible with quantitative scientific modeling. They assessed this approach as producing "great gains" in their understanding of real world problems and scientific research.

The marble game provides a pathway to understanding molecular-level biological systems. Students are also introduced to the basics of computer programming. Finite difference methods are used instead of traditional calculus. This approach more accurately reflects the quantitative and computational methods that are actually used by most practicing engineers and scientists. Students without calculus have been able to successfully complete Module 5 even though the transient diffusion problems they were able to solve (from scratch) are traditionally introduced in advanced calculus-based courses using partial differential equations.

In the marble game approach, students write and then implement their own algorithms starting with a blank Excel spreadsheet. The pedagogical method does not use any computational black boxes. This provides students with an introduction to programming and computer science in an engaging manner that is able to provide an alternate pathway to computational STEM disciplines (PCAST, 2012).

## METHODS

The materials discussed in this paper were originally developed for two courses *Biophysics* (a 300-level course cross-listed in biology, chemistry and physics) and *Physiological Modeling* (a 300-level biology elective). The teaching materials are in the form of a self-contained series of self-study guides (modules) that can be used for both courses. The modules include *optional* calculus sections that are not required for the *Physiological Modeling* course. The pedagogical framework is not calculus-based in the sense of introductory physics courses and all of the materials are accessible to students without calculus. The optional calculus sections make a connection between the finite difference approach used in the modules and traditional calculus – for example see Module 3 (circle4.com/biophysics/modules). It is the primary thesis of this paper that these introductory materials should be placed at the beginning of the quantitative curriculum prior to (or at the same level as) general biology and chemistry.

### Module 1 – The Marble Game

As shown in Figure 1, the marble game has 10 marbles distributed between two boxes. The marbles jump between boxes (in both directions) with a rate constant $k$ because of random thermal motion. This physical system is *simulated* in the marble game by using a ten-sided die to decide which marble next jumps to the other box as a result of the random Brownian motion. The amount of physical time $\Delta t$ that elapses between turns is determined by the rate constant $k$ and the total number $N$ of marbles in the game. The number of marbles in box 1 $N_1$ varies randomly at each turn of the game according to the following rule:

*Marble Game Rule:* Roll the ten-sided die. If you rolled a number *less than or equal to* $N_1$, then move a marble from box 1 → 2, otherwise move a marble from box 2 → 1.

Despite its apparent simplicity, the marble game can be used as a foundation for a conceptual framework that can be applied across the STEM disciplines as outlined below (Ghosh et al., 2006).[2]

*Brownian Motion* The basic move in the marble game is the *unbiased* jump of a marble from one box to an adjacent box (Figure 2) (the jump is unbiased because the jump rate $k$ is the same in both directions). The marble represents one of the molecules that we're interested in (e.g. $O_2$) and the boxes represent two tiny regions. In the simplest case, the two boxes could represent two (essentially identical) adjacent regions of intracellular fluid (cytosol) or they could represent two different regions separated by a membrane (e.g. interstitial fluid and cytosol). Because of random thermal motion, the molecules are constantly jostled by their neighbors. The net result is a passive process that jiggles molecules randomly in all directions. This Brownian motion is the physical mechanism for the jumps between the boxes in the marble game.

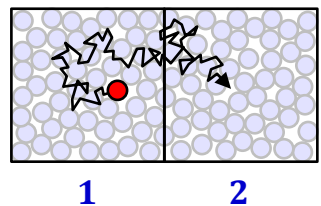

**1**   **2**

Figure 2. Schematic representation of the physical basis for the marble game. Because of random thermal motion, all the molecules are jiggled around constantly. If we focus on a single tagged molecule, this jiggling produces Brownian motion. The black line is a "snail trail" of one possible trajectory that moves the molecule from box 1 → 2. The net effect of this trajectory is that the molecule jumps from box 1 → 2 during a short period of time $\Delta t$.

In Module 1 (Nelson, 2011b), students are guided through implementing the marble game in an Excel spreadsheet. Students discover that equilibrium is a dynamic process and that random Brownian motion is responsible for Fick's (first) law of diffusion. Fick's law is then used to explain the enhancing roles of hemoglobin and myoglobin in the $O_2$ cascade, and why glycolysis enhances glucose transport into cells.

A class activity outlined in the instructor guide to Module 1 (Nelson, 2011b) allows students to actually play with a physical representation of the marble game using a ten-sided die, ten marbles and two boxes. Each student plays their own copy of the game by rolling a die and moving the marbles by hand. To test that students really understand the marble game, each student rolls the die and then calls out the number on their die and their new value for $N_1$ (the number of marbles in box 1). The whole class is responsible for checking that every student used the rule correctly. These results can then be plotted in Excel and the (approximate) *ensemble average* (class average) can also be plotted.

### Module 2 – Algorithms

In Module 2 (Nelson, 2011c) students learn how to assign physical parameters to the marble game and to write a formal kMC algorithm to plan how to implement it in an Excel spreadsheet. This process provides a procedural framework for the remainder of the modules. Students then investigate how the marble game properties depend on system parameters (e.g. $N$ and $k$). By adjusting these parameters, the marble game simulation can be used to model physiological systems ranging from the size of a single molecule up to the entire human body. In the final part of the module, students develop a kMC sim of drug elimination and graphically compare its predictions with the one-compartment pharmacokinetic model and experimental results for elimination of acetaminophen (TYLENOL®).

### Module 3 – Finite Difference Methods

In Module 3 (circle4.com/biophysics/modules/) finite difference (FD) methods are introduced using the marble game as an example (Figure 3). In the FD formulation $N_1$ and $N_2$ now represent the *ensemble average* state of the system at a particular time. Hence, the ensemble average rate of jumps from box 1 → 2 is $N_1 k$, i.e. the number of marbles in box 1 multiplied by the rate at which each one jumps. The arrows in Figure 3 indicate these ensemble average (unidirectional)

---

[2] The marble game model is a kMC variant of Ehrenfest's "dog-flea" model introduced in 1907 (Ghosh *et al.*, 2006).

Peter Hugo Nelson

rates. During a short time interval $\delta t$, the net change in the number in box 1 $\delta N_1$ is indicated by the sum of the two arrows touching box 1. By inspecting the FD diagram (Figure 3), students can write their prediction for how the ensemble average changes with time

$$\delta N_1 = -N_1 k \delta t + N_2 k \delta t \qquad (1)$$

which can be rearranged as

$$\delta N_1 = k \delta t (N_2 - N_1) \qquad (2)$$

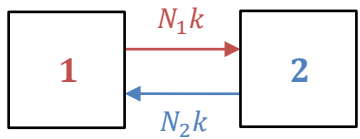

Figure 3. Finite difference (FD) diagram of the marble game. Arrows indicate the average rate at which marbles jump from box to box. The (unidirectional) rates are proportional to the number of marbles in the originating box and the jump rate constant.

Equation (2) is the FD equation for the marble game. By implementing this FD method in a spreadsheet, students are able to **test the hypothesis** that the FD model **predicts** the ensemble average behavior of their previously developed kMC sim. Figure 4 shows an example of the type of graph that students are expected to produce. In this graph, $x_1 = N_1/N$ is the fraction of marbles in box 1. The kMC sim data are "live", so that every time students press the F9 key, the kMC series changes to a statistically independent sample from the ensemble. By answering a series of directed questions, students then discover how system properties vary with time $t$ and system parameters (total number of marbles $N$, the jump rate constant $k$ and the initial fraction of marbles in box 1).

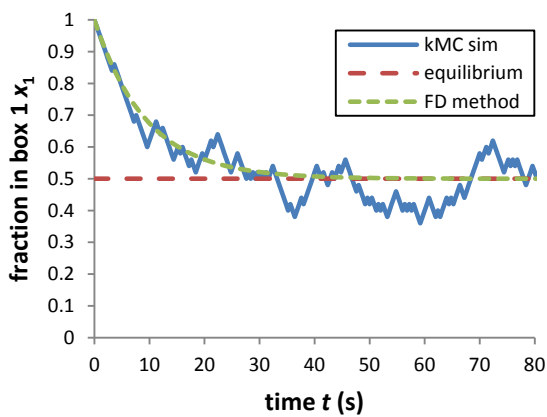

Figure 4. Excel 2010 chart showing the approach to equilibrium of a typical marble game. The graph is a visual comparison of a kMC simulation with the FD model prediction for a system with $N = 50$ and $k = 0.05 \text{ s}^{-1}$.

*Thermodynamics from Kinetics* The general condition for steady-state is that the ensemble average properties of the system do not change with time. Hence, by setting the change $\delta N_1$ in equation (2) to zero, students can derive the equilibrium (thermodynamic) state of the marble game, i.e. $x_1 = 1/2$ at equilibrium (independent of the starting configuration or the jump rate constant or number of marbles $N$ of the particular system).

The FD method is then used to predict the behavior of a normalized order parameter $u$. An analytical equation for $u(t)$ is also presented. Students graphically compare the predictions of the FD model with the analytical equation for various timestep sizes in the FD method and then discover that the accuracy and stability of the FD method depends on the size of the timestep. In an *optional* calculus section, students see how the analytical equation can be derived from the FD formulation.

The FD method is then applied to blood plasma oxygenation. Students discover that the concentration of oxygen dissolved in blood plasma reaches equilibrium with the gas in the lungs within milliseconds of reaching an alveolar capillary. The equilibrium amount of dissolved oxygen is predicted by the FD model and students discover that this model explains the origin of using oxygen partial pressure $P_{O_2}$ as a measure of oxygen concentration (chemical potential) in blood plasma.

## *Complete Module List – Biophysics*

Modules 1-3 provide the conceptual (and procedural) foundation for the rest of the modules. Below is the full list of modules that were used to deliver the course content in *Biophysics* (Spring 2012). Module 4 (https://www.mededportal.org/publication/8081) is an introduction to data analysis and the rest of the modules present more "advanced" topics that may be addressed using the marble game approach.

1. Introduction - marble game
2. Algorithms and pain relief
3. Finite difference method and oxygen
4. Model validation and penicillin
5. Diffusion - spread it around
6. Saturation and least-squares fits
7. Newtonian dynamics
8. Equilibrium distributions – marble game redux
9. Kinetics of the marble game – Poisson processes
10. Molecular dynamics
11. Membrane voltage – electrical polarization
12. Ion channels – voltage-dependent permeases

## *Osmosis (Final Exam Question)*

This section of the paper serves double duty. 1) It provides a concrete example of how the marble game approach can be applied to more advanced topics; and 2) it outlines the final exam question that was used to evaluate whether students could apply their model development skills to a situation that they had not seen before. The question is "Can students apply the modeling process they learned to a novel problem under exam conditions?"

Students were presented with a marble game model of water transport across biological membranes (osmosis) (Figure 5). The importance of this critical factor in physiology was motivated by an extreme example – the woman who died in Sacramento after a water drinking contest called "Hold Your Wee for a Wii". The cause of death was electrolyte and internal pressure imbalances caused by osmosis. Using the approach they learned in the modules, students developed a *novel* marble game model of osmosis and then discover how it can explain the large pressures that result from placing tissues in contact with pure water. In this model the concentration of water in each box is designated $c_1$ and $c_2$. For pure water $c_2 = [H_2O]_{pure} = 55.6 \text{ mol/L}$. If solutes are added to pure water, this "dilutes" the water and the water concentration goes down. The (effective) concentration of water in normal tissue is about $c_1 = 55.3$ mol/L, corresponding to a solute osmolarity of $55.6 - 55.3 = 0.3$ Osm/L.

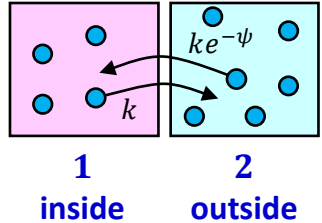

Figure 5. Pressurized marble game. The inside box is at a pressure $\Delta P$ higher than the outer box. The parameter $\psi$ is given by equation (5).

The lipid bilayers that form biological membranes are virtually impermeable to water molecules. As a result, most biological membranes contain water channel proteins (aquaporins) that allow *only* water molecules to pass (Zhu et al., 2004). When a water molecule moves from low to high pressure, mechanical work must be done to push it against the pressure difference. Students were reminded that in Physics I they learned that mechanical work is force times distance $W = Fd$. If that mechanical work was done against a fluid, then the work done is $W = \Delta P \delta V$, which is (hydrostatic) pressure difference $\Delta P$ times the small volume $\delta V$ moved (Figure 6).

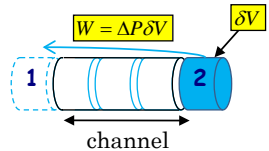

Figure 6. Schematic diagram of an aquaporin (water channel). The net effect of an inward jump is the transfer of a volume of water $\delta V$ (the volume of one water molecule) from box $2 \to 1$. This volume of water moves through a pressure difference $\Delta P$.

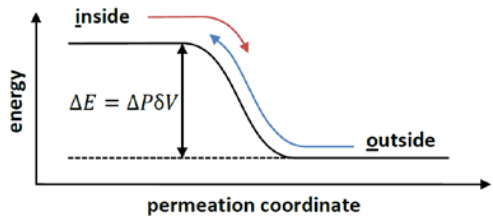

Figure 7. Simplified schematic energy diagram of a water (H₂O) molecule passing through an aquaporin. The diagram shows a situation where the (hydrostatic) pressure is higher inside the cell by an amount $\Delta P$. The water molecule has a volume $\delta V$ and the positive membrane pressure $\Delta P$ raises the water molecule's energy by $\Delta E = \Delta P \delta V$ when it moves from outside to inside.

The schematic energy diagram in Figure 7 shows a simplified diagram of how the water energy changes during the net transfer of a single water molecule through the channel.

$$W = \Delta E = \Delta P \delta V \quad (3)$$

where $\delta V$ is the volume of one water molecule. Equation (3) represents the mechanical work that has to be done (by random thermal motion) in moving a water molecule from outside to inside the cell through the positive pressure difference $\Delta P$.

In an analogous manner to Module 11, the red (left-to-right) arrow represents downward jumps that have a rate constant $k$ for any (positive) pressure difference $\Delta P$. However, because the reverse step in Figure 7 the blue (right-to-left) arrow represents an upward jump with an inward jump rate constant of

$$k_i = k e^{-\psi} \quad (4)$$

where we have defined a dimensionless energy difference

$$\psi = \frac{\Delta P \delta V}{k_B T} \quad (5)$$

which is the ratio of the mechanical work done (energy difference $\Delta E$) to the thermal energy $k_B T$ (Boltzmann constant times absolute temperature). $e^{-\psi}$ is a pressure Boltzmann factor representing the fraction of water molecules attempting to move from box $2 \to 1$ that have enough energy to push another water molecule over the energy barrier (of size $\Delta E = \Delta P \delta V$).

In the question, students were reminded that the energy diagram (Figure 7) is similar to the energy diagrams they saw in Module 11. Students were also given the value of $\delta V = 2.989 \times 10^{-29}$ m³ for the volume occupied by a single water molecule.

Students were then asked to answer:

*Q.1 (i)* When the transmembrane pressure difference $\Delta P$ is zero, what does equation (4) simplify to? *(ii)* How does this zero pressure difference situation compare with the original marble game?

*Q.2* At a normal body temperature of $T = 310$ K, *(i) calculate* the corresponding value of $\psi$ for a pressure difference of $\Delta P = 1$ atm. *(ii)* What fraction of uphill jumps are successful? *Hint:* You will have to be *very careful* with the units you use.

*Q.3 Write* a kinetic Monte Carlo (kMC) algorithm to *simulate* the two-box system. (Students were given values for the required physical parameters $\Delta P, k, T$, etc...)

*Q.4 Draw* a properly labeled FD diagram of the system.

*Reminder:* To get full credit in the following questions, you must show all your working in the manner outlined in Module 2 and Module 3.

*Q.5 Show that* the finite difference equation for change in $c_1$ during a short time $\delta t$ is given by equation (6).

$$\delta c_1 = k(c_2 e^{-\psi} - c_1) \delta t \quad (6)$$

*Q.6* By using the condition for equilibrium, *show that* the equilibrium ratio of concentrations is given by equation (7)

$$e^\psi = \frac{c_2}{c_1} \quad (7)$$

Peter Hugo Nelson

*Q.7* The (effective) concentration of water in normal tissue is about $c_1 = 55.3$ mol/L. Using equation (7) *calculate* the equilibrium pressure difference between tissue and pure water.

*Q.8* Your answer to Q.7 should have been about 800 kPa. *Convert* 800 kPa into atmospheres (atm) and pounds per square inch (psi).

*Q.9 Compare* your answer to Q.8 with 32 psi, which is the recommended pressure for the tires of a small car.

The use of the "show that" style of questions allows students who were unable to successfully derive equation (7) to nevertheless be able to answer the questions that follow. A similar style of questions is used in the modules. A model answer for Q.4 is shown in Figure 8.

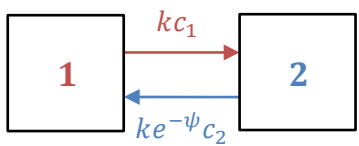

Figure 8. Model answer to final exam *Q.4* – an FD diagram of the pressurized marble game representation of osmosis. The parameter $\psi$ is given by equation (5).

*A Kinetic Model of Osmosis* It is common for undergraduate physiology instructors to describe osmosis as the "diffusion" of water between two compartments. The (thermodynamic) driving force for the motion of water from low to high ***solute*** concentration is explained as being caused by the "diffusion" of water from high to low ***water*** concentration (Baierlein, 2001; Zhu et al., 2004). It is a significant advantage of the marble game approach that it provides a quantitative model consistent with this traditional explanation of osmosis.[3] However, there is an apparent controversy. The analogy between osmosis and diffusion has recently been characterized as a misconception that fails to make correct quantitative predictions (Kramer and Myers, 2012). Hence, it is important to show that the marble game framework produces a quantitative model that is consistent with well-known thermodynamic principles.

In answer to Q.7, students were expected to derive equation (8)

$$\Delta P = \frac{k_B T}{\delta V} \ln \frac{c_2}{c_1} \quad (8)$$

This equation is the osmotic analogue of the Nernst equation (equation (12)). To date, I have not been able to find equation (8) in the research or pedagogical literature. If the concentration ratio is replaced with the activity ratio and the activity of pure water is set to $a_2 = 1$, equation (8) reduces to equation (9.3.13) in Truskey *et al.* (2009) that was derived based on traditional thermodynamic arguments. As they show, this general equation reduces to the more familiar van't Hoff equation for the osmotic pressure $\pi$ in the limit of a dilute solution

$$\pi \approx c_s RT \quad (9)$$

where $c_s$ is the solute concentration and $R$ is the gas constant. The van't Hoff equation is a good approximation to the more thermodynamically accurate equation (8) over the physiologically relevant range. For example, at an osmolarity of 1 Osm/L (the approximate osmolarity of sea water) the difference between them is less than 1%. Hence, the marble game approach provides an intuitive model of osmosis that is thermodynamically sound and reduces to the standard equation (9), but also provides a kinetic (diffusive) explanation for osmosis within the conceptual framework provided by the marble game. The pressure Boltzmann factor in equation (4) provides for the required "rectification of Brownian motion" through the channel (Kramer and Myers, 2012; Nelson, 2007).

---

[3] This conceptual model can be made rigorous by adopting the "tendency to diffuse" or "rate of change" meaning of chemical potential (Baierlein, 2001). If properly constructed, the jumps between boxes in the marble game (Helmholtz ensemble) representation of osmosis will reach equilibrium when the chemical potential of the water molecules is the same in each box, similar to a more realistic simulation of an aquaporin (Zhu *et al.*, 2004). A similar approach is also commonly used in physiology to describe the chemical potential of dissolved oxygen in terms of a "partial pressure" of $O_2$ dissolved in blood plasma (Module 3).

The marble game model of osmosis (Figure 5) is directly analogous to that used in Module 11 for permeation of ions across lipid bilayer membranes through ion channels. The permeation mechanism (Figure 6 for osmosis) can be replaced with FIG. 1 of (Nelson, 2011a) for ion channels. The energy diagram (Figure 7) is the same for ion permeation, except that the mechanical work $W$ is replaced with electrical work $W_{elec}$.

$$W_{elec} = \Delta E_{elec} = qV_{mem} \quad (10)$$

In equation (10) $q$ is the charge of the ion (e.g. $K^+$) and $V_{mem}$ is the (positive) membrane voltage. In that case, the system is also characterized by a dimensionless energy difference

$$\psi = \frac{qV_{mem}}{k_B T} \quad (11)$$

and $e^{-\psi}$ is an electrical Boltzmann factor.

Electrical equilibrium is again defined by equation (7) (with $\psi$ now defined by equation (11) and with $c_1$ and $c_2$ representing the permeant ion concentrations in boxes 1 and 2 respectively). An algebraic rearrangement of equation (7) then results in the Nernst equation.

$$V_{mem} = \frac{k_B T}{q} \ln \frac{c_2}{c_1} \quad (12)$$

This kinetic model of ion permeation (analogous to the osmosis model discussed above) was first described by Armstrong for an ideally thin membrane (Armstrong, 2003). This thin membrane model can be extended to concerted permeation mechanisms such as the one illustrated in Figures 6 and 7 using a recent theoretical framework (Nelson, 2003; Nelson, 2011a) that explains ion channels as voltage-dependent permeases.

*Teaching Method*

The teaching method used for the *Biophysics* course (Spring 2012) was a "flipped classroom" or "inverted

classroom"(Gannod et al., 2008; Lage et al., 2000) – an extreme version of the "learn before lecture" method, which has been shown to improve student performance (Moravec et al., 2010). In most flipped classrooms, students watch video lectures (e.g. from www.khanacademy.org (MacIsaac, 2011)) at home and then work on problem-solving or "homework" in the classroom. *Biophysics* used a modified flipped classroom method. All course content was presented to students in the form of module PDF files that were posted online before class. The modules are self-contained study guides that include inline questions intended to engage students in discovery-based activities (Modules 1, 2 and 3 are available at circle4.com/biophysics). Specific questions contained in the modules were assigned (in advance) for each class for a small amount of credit. Students were required to make an initial attempt at these questions before each class period and then to submit their draft answers before class. These preliminary answers were then assigned full "attempt credit" if it appeared that the student had made a good-faith attempt to answer the questions. Use of this attempt credit greatly enhanced student preparation for the in-class problem solving activities. I have taught *Biophysics* and *Physiological Modeling* for a combined total of 14 semesters and this innovation was by far the most successful strategy for getting students to come to class well prepared.

## *Evaluation methods*

The primary formative evaluation of these materials has been direct interaction with students working through the module questions in class. The completed module answers then provided more anecdotal evidence for the effectiveness of the modules. This evidence was further supported by two midterm tests.

The evidence presented below is based on results from *Biophysics* (Spring 2012) ($N = 13$ students). The first is a short-answer pre-post test of key concepts (the pretest was conducted on the first day of class and the posttest was included in the final exam). Data from student answers to the osmosis final exam question are also presented. Finally, a Student Assessment of their Learning Gains (SALG) survey (www.salgsite.org) (Seymour, 2000) was administered after the final exam, but before final grades were posted. ***All student comments were extracted verbatim and in their entirety*** from the open-text response portion of the SALG survey. The results are presented below in a similar manner to Casem (2006).

## RESULTS

The following results are preliminary, but indicate the effectiveness of the approach for teaching students a coherent conceptual framework using self-contained modules.

## *Pre-Post Test*

As evidence that students learned the concepts presented in the modules, a short answer pre-post test was conducted using the following questions.

1. What is equilibrium?
2. What causes diffusion?
3. What is the Nernst equation?
4. What is the molecular *cause* of Fick's first law? In other words, why do molecules move (on average) from high to low concentration?
5. Why does the Nernst potential depend on the logarithm of concentration?

For the purposes of evaluation, student answers were assigned a score based on whether they volunteered a correct concept (Figure 9). For the equilibrium question, full credit was assigned if the student offered that the forward and reverse rates are equal (or otherwise indicated that equilibrium is a dynamic process). If students answered that the concentrations were balanced, or that the reaction was complete, their answer was assigned half points (no indication of a dynamic process). For the diffusion question, "concentration gradient" or "difference in concentration" was correct. In the pre-test no students identified "Brownian motion" whereas 4/13 students included this in the post-test answers. For the Nernst equation question, the mathematical equation (12) or a verbal description of it was correct. "A thermodynamic equation" received half points. For question 4, the correct concept was "Brownian motion" or "random motion". Common misconceptions such as "molecules go where they are needed.", "molecules do not want to sit next to each other and therefore spread out as evenly as possible", "going from high to low gives the molecules more space" and "there is less force on the molecules when they are less crowded" earned zero points. For question 5, the correct concept was either "the Boltzmann factor" or "entropy". The results for all of these questions show substantial improvement, indicating the effectiveness of the approach.

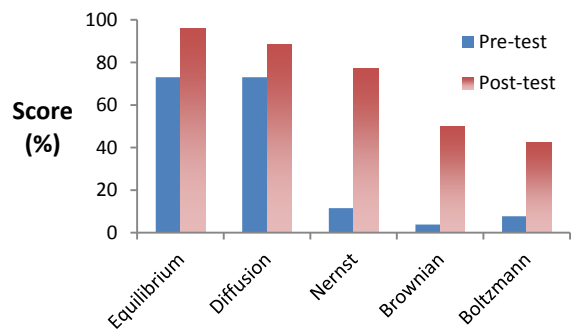

Figure 9. Pre-post test data for *Biophysics* (Spring 2012). The average student score ($N = 13$) for each conceptual question is shown as a percentage for each open-ended conceptual question.

## *Osmosis – Final Exam Question*

In the *Biophysics* final exam, students were expected to develop a *novel* model of osmosis (from scratch) as 30% of a two-hour closed-book-with-cheat-sheet final exam. Osmosis is a concept that is well-known in the pedagogy literature as being a difficult subject that involves many student misconceptions (Ben-Sasson and Grover, 2003; Kramer and Myers, 2012; Meir et al., 2005; Odom, 1995; Tekkaya, 2003). This topic was not previously discussed in *Biophysics*. As part of this final exam question, students were asked to: write a kMC algorithm; draw a properly labeled finite difference diagram; derive a finite difference equation; derive the equation for osmotic equilibrium; solve that equation for the osmotic pressure; calculate a numerical osmotic pressure value for tissue; and then perform a unit conversion.

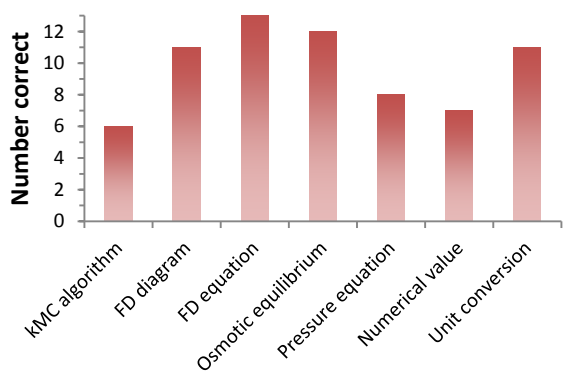

Figure 10. Number of correct student answers ($N = 13$) for 7 parts of the osmosis final exam question.

Poor results on the kMC algorithm (Figure 10) reflect the fact that writing a complete kMC algorithm (using the Metropolis algorithm (Metropolis et al., 1953)) is a difficult (high Bloom order - creative) process. This was compounded by the fact that the question format presents a physical situation to students and they have to distinguish between two representations kMC (stochastic) followed immediately by FD (deterministic) – many students (5/13) answered the algorithm sub-question as if the whole question was about the FD modeling technique. However, results for the next three questions are extremely encouraging. They show that 11/13 students were able to correctly construct an FD diagram from the marble game representation, 13/13 were able to correctly derive the FD equation and 12/13 were able to correctly derive the thermodynamic equation for osmosis. To summarize, at this point, students (through a series of directed questions) have correctly derived a *novel* theoretical model for osmosis – equation (7), based on a marble game representation of the problem (Figures 5-7). If students understand how they got to this point, then they have a working kinetic and thermodynamic model of osmosis that is based on the marble game conceptual framework.

The remaining parts of the question are similar in level to what might be asked in a traditional introductory (algebra-based) physics course – solving for something in the exponent of an exponential function and then solving two simultaneous equations for the pressure equation. The numerical value question is basically "plug and chug" at the introductory physics level. The unit conversion question is at the high school level. All of these introductory level questions had lower scores than the questions relating to student application of the marble game conceptual framework to a novel situation.

## SALG Survey

*Modules are Self-contained* Student comments (below) confirm that that the subject matter of the course was presented in the form of self-contained self-study modules. Student responses highlight both the benefits and the disadvantages of this instructional approach.

"I am not sure about the resources we used. All we used were the modules."

"This class was not taught at all. It was self-taught. I wish there was more lecture. I like it when Dr. Nelson lectures :("

"The modules allowed for a very self-directed learning style which I feel allowed me to absorb the material better."

"It was very hands on and I liked that"

"The self-directed approach made it easier for me to learn."

"Completing the modules helped my learning tremendously. Everything else during class was not really useful since we didn't really do anything in class."

"We all collaborated and helped each other understand what to do for the modules."

"I liked the way we made small groups and everybody helped each other with graphs."

"WE worked together and figured out things we couldnt have alone. It took a lot less time too when we helped each other"

*Study and Skills Development* Because of the flipped classroom environment, students were forced to actually read the modules to be able to complete the daily and weekly assignments. This served the purpose of testing whether the materials could be used as a stand-alone resource. However, a flipped classroom is probably not the optimal method (Baldwin, 2009; Bybee and Westbrook, 2006; Felder, 1998; Froyd, 2008; PCAST, 2012; Tanner et al., 2003). The comments below illustrate student reflections on how their study habits and skills changed as a result of the course.

"Doing algorithms over and over again! Of course having parts of the modules due before each class really helps to keep us on our toes and not to leave the whole module to do last minute."

"Reading the modules is a bit tough, it takes two or three times to read it over before the message really sinks in. The skills I have acquired I know will really help me with analyzing my research data."

"I had to learn how to use my time wisely."

"More critical reading skills."

"A lot of skills in validating a model and statistics."

"This was the most independent studying I did at BU, so it increased by confidence in independently studying."

"don't leave the work to the last second anymore. keep up with a steady amount of work to make sure it gets done"

"Reading is key. Procrastination is very bad. Understanding is more important than memorizing."

*Interdisciplinary Conceptual Framework* Student responses below reflect their understanding that the marble game approach can provide a (well-organized) unifying conceptual framework leading to an interdisciplinary learning progression (Klymkowsky and Cooper, 2012; Schwarz et al., 2009). The first comment reflects student assessment that the first three modules could work at the introductory level. The last comment summarizes the transformative effect on one student's worldview.

"The module on after another was a bit intense. The first two or three modules definitely could be implemented into University physics I/II so that when students start biophysics they can pick right up from there. This may leave time in the semester to actually discuss the modules so that students can fully understand the "about what you discovered" sections."

"Everything integrated really well together"

"very organized and learned great stuff and gained lots of knowledge"

"I have understood a lot of stuff that I was confused about in other classes."

"…I would absolutely take another class in Biophysics. This class teaches you the actual dynamics of what you learn in physio. It is a very interesting subject."

"My Excel knowledge and techniques has greatly improved. The different physiological modeling done in the modules has helped me to understand what is actually going on in our bodies. It is a very unique way to teach physics and physio."

"Many different topics can be used to present in the same way, and many topics in the sciences are very interconnected."

"I was able to develop a stronger understanding of the fundamental basis for a lot of larger concepts. This significantly enhanced my understanding of those larger concepts."

"we got to know how things works actually rather than reading int the books"

"The algorithms have helped me create a new method of viewing relationships and remembering data"

"Excel techniques for sure! But the algorithms help to organize the thought process of trying to do step by step equations. That thought process and organization I will definitely take with me."

"Algorithms and the style of thinking by breaking everything down."

"i found it interesting how biological systems and patterns could be predicted and observed using algorithms and physics"

"It will help me a great deal in future kinetics classes and situations"

"I never knew models could apply to virtually any process given the right algorithm. This is crazy!!!"

*Scientific Research Applied to Real World* Students assessed that they had made great gains in their ability to perform real (quantitative modeling) scientific research (Table 1). Students perceived that their learning gains were greater for real scientific research and real world issues than for other classes. Student's written comments (below) also support the assertion that students perceive that the materials relate to real world issues and scientific research. Students also identified specific methods used the course that they perceived as relating to real scientific research.

**Table 1. Results for the SALG question:**
**As a result of your work in this class, what GAINS DID YOU MAKE in your UNDERSTANDING of each of the following?**

| Question | Rating |
| --- | --- |
| How ideas from this class relate to ideas encountered in other classes within this subject area | 4.2 |
| How ideas from this class relate to ideas encountered in classes outside of this subject area | 3.7 |
| How ideas from this class relate to real scientific research | 4.5 |
| How studying this subject area helps people address real world issues | 4.5 |

"The intense/careful reading of the modules has definitely helped me to read scientific material better. Excel techniques have definitely improved as well and can be applied to data analysis."

"i can make an algorithm and implement it into excel to determine what the actual data would do compared to the theoretical model used to fit the data"

"I can work with excel really well as a result of this class and understand the importance of algorithms to scientific models"

"Biophysics323 is a unique research-based class that teaches you real-life research skills in building models, testing them, and constructing publication-quality graphs on Excel."

"This class taught me methods that are very useful in my research. In fact, I would guess that there is no other class quite like this in its unique approach to teaching you how to be a scientific researcher and biophysicist."

*Student Assessment of Learning Gains by Module* In the SALG survey, students rated "Graded assignments (overall) in this class" as being of great help to their learning (mean 4.6 with 8/12 reporting "5:great help"). Rankings for the individual modules appear to fall into two categories (Figure 11). Modules 1-4 had a mean of 4.8 with (an average of) 10/12 students reporting "5:great help" to their learning. The later Modules 6-12 were generally ranked lower (mean 3.8 with 4/12 students reporting "5:great help" to their learning and with 4/12 students reporting "4:much help" to their learning. These two categories correlate with the state of development of the modules. Modules 1-4 are in close to final form, whereas the rest of the modules are still under development. Consistent with this trend, the two lowest rated modules 7 and 10 were presented for the first time in *Biophysics* (Spring 2012). The module development cycle that was used is based upon on constant formative assessment by students.

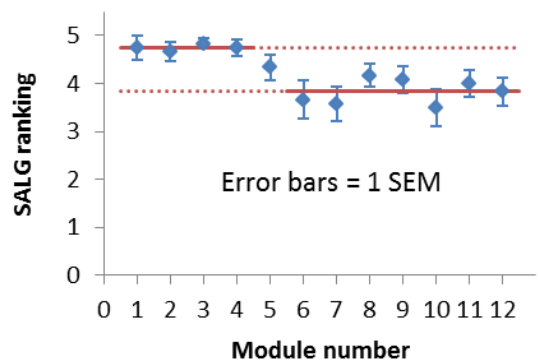

Figure 11. Student evaluations of the course material by module. SALG ranking is on a 5-point scale and the errors bars indicate one standard error of the mean value (SEM). The distribution of rankings is bimodal with students ranking the first 4 modules higher than the last 7, perhaps reflecting the state of development of the modules (see text).

## DISCUSSION

The marble game provides a new starting point for teaching quantitative modeling across the STEM disciplines. Using a series of self-contained self-study guides (the modules), students were able to implement the marble game in an Excel spreadsheet. Students were then able to engage in a process of directed scientific discovery that they identified as strongly relating to real world problems and real scientific research. The conceptual framework developed in the modules is intrinsically interdisciplinary and the study skills learned by students are foundational to success in all of the STEM disciplines.

Despite the simplicity of the marble game (one repeated rule), it provides a computational framework consistent with the chemical master equation (Gillespie, 1977; Nelson et al., 1991). This framework is useful for modeling a wide range of physical processes. Because students do not know what to expect from the marble game, it provides them with an inquiry-based introductory experience based on interpreting graphical data to discover how a system works. Students

Peter Hugo Nelson

discover for themselves that this randomness is the reason for Fick's (first) law of diffusion and results in a dynamic equilibrium (Garvin-Doxas and Klymkowsky, 2008).

In Module 2 students learn the basics of algorithmic thinking. The results of the SALG survey show that students understood the importance of this new conceptual paradigm. The skill of learning how to write an error-free algorithm is probably the *most difficult skill* for many students in these modules. The use of Excel mitigates this problem because the algorithm is implemented spatially in the cells of a spreadsheet, requiring less formal syntax than a traditional programming language. In addition, most students are already familiar with Excel, which is a significant advantage because students are not intimidated by it (there were no negative student comments about the use of Excel in the SALG survey).

The FD formulation of the marble game (Module 3) introduces students to a simple computational model. The advantage of the marble game's introductory approach is that the mathematical modeling occurs with a system that is already somewhat familiar, but that is still somewhat mysterious. Students then engage in what they assessed themselves as being an authentic scientific experience – testing that the deterministic FD model predicts the ensemble average behavior of the inherently stochastic kMC sim of the marble game. In this manner, students graphically compare two different models of the same system (Figure 4).

Students also learn the utility of different state variables, $N_1$ (number) is extensive, $c_1$ (concentration) is intensive and $u$ is a normalized order parameter for the system. This last system variable is extremely useful in making the connection between the FD method and a traditional calculus formulation of the marble game system. Thus, the marble game also provides excellent motivation for the application of traditional calculus techniques to science; technology and engineering – using a framework that is accessible to students with no prior exposure to calculus.

As students assessed, the conceptual framework provided by the marble game can be applied to many problems in STEM. The basic game is directly applicable to molecular transport in biology under a wide variety of circumstances, diffusion, osmosis, ion permeation, interfacial transport in the lungs, on the skin and at epithelia throughout the body and the kMC approach can produce realistic models of genetic and population dynamics. The marble game is also isomorphic with a first-order reversible chemical reaction with the jump rate constant replaced with a chemical reaction rate constant. Hence, the framework developed can be used to introduce kinetic modeling (at the level of general chemistry). The advantage of the marble game (over traditional chemical kinetics) is that the rate constant characterizes a simple transport process that can be easily visualized.

## *Discovery-Based Physics*

As outlined in the introduction, it is the central thesis of this paper that the marble game is an ideal starting point for mathematical and computational scientific modeling across the STEM disciplines (including physics). In the *Matter and Interactions* approach (Caballero et al., 2012; Chabay and Sherwood, 2008) the starting model is of a particle moving through three dimensional space. While this may be a good starting point for well-prepared physics and engineering majors (who have previously succeeded at algebra-based physics in high school), it is probably not a good starting point for many life science majors (who have either not taken physics before or who have had conceptual difficulties with the traditional approach). The issue is twofold.

1) Newtonian mechanics is a subject with which students have a great deal of real world experience, but they often have misconceptions.

2) Newtonian mechanics is based upon $\vec{F} = m\vec{a}$, which appears simple, except for the fact that acceleration $\vec{a}$ is the rate of range (first time derivative) of velocity, or the second time derivative of position. In addition, all the variables are vector quantities. This means that students must have a good grasp of trigonometry and geometry (including vector addition and subtraction) before they can be expected to understand the quantitative modeling process. This mathematical complexity (at the beginning of the course) severely compounds the problem of student misconceptions (Caballero et al., 2012; Hake, 1998; Knight, 1995). This is a curricular issue – of what we teach and how it affects student learning (Klymkowsky and Cooper, 2012).

From a mathematical point of view, the marble game is much simpler than Newtonian mechanics, but the computational framework it provides can be used to investigate the surprising mathematical simplicity of dynamics-with-constant-acceleration in a discovery-based computational approach. One way to do this is to postulate a constant acceleration of $a = 22 \frac{\text{mi}}{\text{h}}$/s for a falling object (or for an accelerating Bugatti Veyron car), and then investigate the consequences for the motion using a finite difference approach similar to equations (1) and (2) and a spreadsheet to calculate the changes in speed ($\delta v = a\delta t$) and position ($\delta x = v\delta t$). Using this framework (and a series of directed questions), students can continue their inquiry-based active-learning approach for traditional mechanics topics (Module 7) that can then progress through introductory computational physics (Chabay and Sherwood, 2008) using a discovery-based approach. As an example, the learning progression allows students to input the empirical observation that Hooke's law $F = -kx$ explains springs into their own FD computational framework and then discover that it produces simple harmonic motion (Module 10). This approach can be extended beyond Newtonian mechanics into other areas of physics including RLC circuits and waves.

The material in the introductory sequence (Modules 1-3) is required before the marble game approach can be applied to other STEM areas. This can take a significant amount of time – about 3 weeks of a 3 – 4 credit-hour course, but the benefits are worth that effort, particularly within a multi- or inter-disciplinary context. In addition, student enthusiasm for this approach should aid in retaining first-year students in the STEM disciplines and prepare them to succeed in later STEM courses (PCAST, 2012).

Because the materials are self-contained, many students should be able to complete the introductory modules outside of a traditional classroom setting, potentially creating new pathways to the STEM disciplines (PCAST, 2012).

## CONCLUSIONS

Using the marble game, students can be introduced to quantitative scientific modeling in a process of directed scientific discovery. The marble game provides a learning progression that spans the STEM disciplines. It is directly applicable to many important molecular-level concepts in biology such as Brownian motion, diffusion, osmosis and interfacial transport in a manner that integrates the inherent randomness of molecular systems. Because it is isomorphic with a first-order chemical reaction it can also provide a cross-disciplinary introduction to chemical kinetics. Students also develop algorithmic thinking skills that are foundational for computer science and an understanding of the predictive power of differential (FD) models that can motivate the formal study of calculus (PCAST, 2012). Finally, after students have mastered the computation approach introduced using the marble game, they can then use it to apply the scientific method to test the hypothesis that Newtonian mechanics *predicts* projectile motion etc. This computational approach can then be used throughout the traditional physics curriculum.

As shown by the results of the SALG survey, students in *Biophysics* enjoyed using the marble game and they reported great gains in many of the areas of pedagogical need – including appreciating that science is an evidence-based endeavor that relates to the real world. However, *Biophysics* was a course exclusively populated by seniors, most of whom (11/13) were biochemistry and molecular biology majors, which puts the introductory portion of the material in the wrong place in the curriculum. These materials need to be tested at an introductory level.

The modules are self-contained self-study guides that can be implemented using any computing device that is able to run a spreadsheet program. They are ideal for independent study and are thus well-suited to provide alternate pathways to STEM (PCAST, 2012). As the modules require only minimal resources, they might be appropriate for programs targeting under-represented minorities. Additional evidence of their effectiveness will be required to motivate implementation of the proposed curricular reform - to include quantitative biology concepts in the undergraduate curriculum - from the very first class.

## ACKNOWLEDGMENTS

Funding from the NSF TUES Program DUE-0836833 (formally CCLI) is gratefully acknowledged. I would like to thank Robert Hilborn, Michael Klymkowsky, Jaqueline Lynch, Heather Masonjones, Niina Ronkainen, Philip Schreiner and Stefan Winkler for their insightful comments and advice.

Peter Hugo Nelson